\documentclass[hyper]{JHEP3} 
\usepackage{epsfig,amsmath}

\newcommand{\newc}{\newcommand}
\newc\eg{{\it {e.g.}}}	\newc\vs{{\it {vs.}}}	\newc\etal{{\it {et al.}}}
\newc\etc{{\it {etc.}}}	\newc\ie{{\it {i.e.}}}

\newcommand{\charn}{\chi^{-}} 

\newc{\mhalf}{m_{1/2}}      \newc{\mzero}{m_0}

\newc{\pleft}{P_L}      \newc{\pright}{P_R}

\newc{\sckm}{\mbox{\rm sCKM}}  \newc{\sckmzero}{\mbox{\rm sCKM}^{(0)}}

\newcommand\gluino{\widetilde{g}}

\newc\bsgamma{b\rightarrow s\gamma} \newc\bxsgamma{B\rightarrow X_{s}\gamma}
\newc\brbsgamma{BR(B\rightarrow X_s\gamma)}

\newc{\tanb}{\tan\beta}
\newc{\azero}{A_0}
\newc{\at}{A_t} \newc{\abot}{A_b} \newc{\atau}{A_\tau}
\newc{\bmu}{B\mu}           \newc{\sgn}{{\rm sgn}}
\newc{\mone}{M_1}           \newc{\mtwo}{M_2}

\newc{\bino}{\widetilde B}              \newc{\wino}{\widetilde W_3}
\newc{\higgsinob}{{\widetilde H}^0_b}   \newc{\higgsinot}{{\widetilde H}^0_t}

\newc{\mtop}{m_{\rm t}}
\newc{\mbottom}{m_{\rm b}}

\newc{\mw}{m_{\rm W}}
\newc\msusy{M_{\rm SUSY}}
\newc{\mplanck}{M_{\rm P}}

\newc{\mub}{{\mu}_{\rm b}}	  \newc{\muw}{{\mu}_{\rm W}}
\newc{\mususy}{{\mu}_{\rm SUSY}}  \newc{\muzero}{\mu_{\rm 0}}
\newc{\deltaw}{\delta^{W}}	  \newc{\deltah}{\delta^{H}}
\newc{\deltas}{\delta^{S}}	  \newc{\deltacharn}{\delta^{\charn}}
\newc{\deltax}{\delta^{X}}	  
\newc{\deltaneut}{\delta^{\chi^0}}  \newc{\deltagluino}{\delta^{\gluino}}

\newc{\Ci}{C_i}	\newc{\Cip}{C_i^{\prime}}

\newc{\deltadll}{\delta^d_{LL}}	\newc{\deltadlr}{\delta^d_{LR}}
\newc{\deltadrl}{\delta^d_{RL}}	\newc{\deltadrr}{\delta^d_{RR}}

\newc{\abund}{\Omega h^2}
\newc{\abundchi}{\Omega_\chi h^2}
\newc{\rhocrit}{\rho_{crit}}
\newc{\rhochi}{\rho_{\chi}}

\newc{\xf}{x_f}
\newc{\jxf}{J({\xf})}
\newc{\VEV}[1]{\langle #1 \rangle}

\newc{\ra}{\rightarrow}
\newc{\beq}{\begin{equation}}
\newc{\eeq}{\end{equation}}
\newc{\bea}{\begin{eqnarray}}
\newc{\eea}{\end{eqnarray}}

\newcommand\lsim{\mathrel{\rlap{\lower4pt\hbox{\hskip1pt$\sim$}}
    \raise1pt\hbox{$<$}}}
\newcommand\gsim{\mathrel{\rlap{\lower4pt\hbox{\hskip1pt$\sim$}}
    \raise1pt\hbox{$>$}}}

\newcommand{\Rn}[1]{{\uppercase\expandafter{\romannumeral#1}}}

\newcommand{\epsfile}[1]{\relax}
\title{Similarity between Kaluza-Klein and Open-string amplitudes \\
in Diphoton Production }

\author{Piyabut Burikham\\
        Department of Physics, University of Wisconsin,
        1150 University Avenue, Madison, WI 53706\\
        E-mail: \email{piyabut@pheno.physics.wisc.edu}}

\abstract{We calculate the tree-level open-string amplitudes for the scattering of four massless particles with diphoton final states.  These amplitudes are required to reproduce those of standard model at the tree level in the low energy limit.  After low energy stringy corrections, we found that they have similar form to the same processes induced by exchange of the Kaluza-Klein(KK) excitations of graviton in ADD scenario.  Using this similarity, we apply constraints on the KK mass scale $M_D$ to the string scale $M_S$.  The results are consistent with constraints from the 4-fermion scattering, about $0.6-0.9$ TeV. }

\keywords{Extra Large Dimensions, D-branes, Beyond Standard Model}

\begin{document}


\section{ Introduction}\label{sec:intro}

Scatterings in TeV-scale string scenarios have been proved to be phenomenologically viable.  In these models~\cite{add,pes,cor,bhhm,shi,rs}, gravity is naturally weakened due to freedom to propagate into extra dimensional direction in the bulk while standard model particles are identified as open string whose ends are confined to D-branes, usually assumed to be 3 dimensional.  Since string theory requires ten dimensional spacetime and we have experimental limit on size of extra dimension $R<200~\mu$m\cite{gra}, compactification is an inevitable.  Boundary conditions of compactified dimensions discretize momenta perpendicular to brane and give rise to discrete spectrum of Kaluza-Klein(KK) excitations.  Each standard model particle can be assigned to have a corresponding KK tower of states with the same 4 dimensional quantum numbers and masses roughly proportional to the inverse of size of the compactified radii of extra dimensions.  Parallel to the brane, \ie~our matter universe, scattering of standard model particles is calculated as the open-string scattering amplitude\cite{pes,cor,jos,bhhm} with string coupling $g_s$ identified as $g^2_{YM}$.  As we approach higher energy, stringy behaviour of open-string scattering become visible, specifically contribution from string resonances(SR) become significant.  We can calculate deviations from typical standard model amplitudes and put constraints on string scale, $M_S$, using experimental data from particle accelerators\cite{pes,bhhm}.                  

In conventional Kaluza-Klein models where only graviton has KK modes being decomposed into spin-2,1, and 0\cite{kk}, there are corrections to standard model amplitudes due to the exchange of graviton and KK excitations.  Each mode of the KK contribution is suppressed by the Planck mass, $M_{Pl}$.  These KK-states interact only weakly $\sim 1/M^2_{Pl}$ with particles on the brane.  However, eventhough each KK exchange is suppressed by the Planck mass, when we sum over the tower of states from $1/R$ to $M_D$, the total contribution adds up to $\sim 1/M^4_D$ which could be in the range of the TeV-scale without being ruled out by previous experiments. This opens up possibility that there is an unobserved tower of KK-states with mass $\sim 1/R$ spanned in low energy ranging from less than 1 eV upto some cut off scale $M_D$.       

In this paper, we will show that in certain processes; such as diphoton production, there is remarkable similarity between amplitudes from SR and KK exchanges in both the angular distribution and the energy dependence aspects.  Since SR amplitudes are calculated from the scattering of open strings on the brane while the KK amplitudes are extra-dimensional corrections from the bulk components of gravitons, their similarity is therefore something of curious nature.  It also suggests that we might as well find two copies of similar contributions in the form of dimension-8 operator in future colliders for diphoton production processes. 

\section{Open-string amplitudes for diphoton production processes}

Since all the relevant diphoton processes of KK model in ADD scenario have been calculated in ref.~\cite{kkg}, we will calculate only open-string amplitudes for diphoton production.  One of the processes which is of importance in an open-string model is scattering of two photons into two photons(4-photon scattering).  There are two reasons for special interest in this process.  First, the amplitude vanishes at the tree-level in standard QED; it is a 1-loop effect and thus is very weak.  Secondly, since photon is identified with the $U(1)$ sector of any open-string models, its Chan-Paton matrix is always diagonal and unique.  Therefore the Chan-Paton factors(trace of four Chan-Paton matrices) can never be zero.  This results in a universal form for the non-vanishing 4-photon scattering amplitude in open-string models with only one undetermined parameter, the string scale $M_S$.  Constraints on the string scale from 4-photon scattering is therefore a definite and universal condition applicable to every model in braneworld scenario.          

We will consider 4 possible initial 2-particle states that give diphoton final state, namely, $\gamma \gamma, q\bar{q}, gg$ and $\ell \bar{\ell}$.    

\subsection{4-photon scattering}

The open-string tree-level amplitudes for 4-photon scattering can be expressed generically\cite{pes,has,man}(see also Appendix) for each helicity combination as:
\vskip 2mm
\underline{(1). $\gamma_{\alpha}\gamma_{\alpha} \to \gamma_{\alpha}\gamma_{\alpha};\  \alpha=L,R$ }
\begin{eqnarray}
A_{string} & = & ig^2 T\frac{s}{ut}f(s,t,u) 
\label{eq:1}
\end{eqnarray}
where $T\equiv T_{1234}=T_{1324}=T_{1243}$ are Chan-Paton factors and $f(s,t,u)\equiv uS(s,t)+sS(t,u)+tS(u,s)$; is $(s,t,u)$-symmetric function which vanishes in the low energy limit.  
\vskip 2mm
\underline{(2). $\gamma_{\alpha}\gamma_{\beta} \to \gamma_{\alpha}\gamma_{\beta};\  \alpha,\beta=L,R;\ \alpha\neq\beta$ }
\begin{eqnarray}
A_{string} & = & ig^2 T\frac{t}{us}f(s,t,u) 
\label{eq:2}
\end{eqnarray}
where $T\equiv T_{1234}=T_{1324}=T_{1243}$ are again Chan-Paton factors.  For $\gamma_{\alpha}\gamma_{\beta} \to \gamma_{\beta}\gamma_{\alpha}(\alpha\neq\beta)$, we simply exchange $t\leftrightarrow u$ in the second case.  Note that Chan-Paton factors for all helicity combinations are the same because the trace of a product of diagonal matrices does not depend on the ordering.  Due to its Abelian nature, the massless photon is always represented by diagonal commuting(with respect to each other) matrices.  In the low energy approximation ($s,t,u \ll M_{S}^2;S(s,t) \approx 1-\pi^2 st/6M_{S}^4, f(s,t,u)\approx -\pi^2 stu/2M_{S}^4$), these amplitudes give     
\begin{eqnarray}
\bar{\Sigma}|A_{string}|^2 & = & \frac{1}{8}(\pi^2Tg^2)^2\frac{s^4+t^4+u^4}{M_{S}^8}
\label{eq:3}
\end{eqnarray}
We will identify the coupling $g$ with the QED coupling, $e$, since photon is associated with QED by definition.  

\subsection{Diphoton Production at the Tevatron}

There are contributions from quark-antiquark and gluon-gluon initial states.  First we consider quark-antiquark amplitudes:

\vskip 2mm
\underline{(1). $q_{\alpha}\bar{q}_{\beta} \to \gamma_{\alpha}\gamma_{\beta};\  \alpha,\beta=L,R;\ \alpha\neq\beta=L,R$ }
\begin{eqnarray}
A_{string} & = & ig^2 \left[T_{1234}\frac{\sqrt{ut}}{s}+T_{1324}\sqrt{\frac{t}{u}}+T_{1243}\frac{t}{s}\sqrt{\frac{t}{u}} \right] 
\label{eq:4} \\
A_{QED}    & = & 2ie^2Q_{q}^2\sqrt{\frac{t}{u}}  
\end{eqnarray}
Matching with $g=e$ gives;
\begin{eqnarray}
T_{1234} & = & T_{1243}\equiv T \\
T_{1324} & = & T + 2Q_{q}^2 \\
A_{string} & = & ie^2 T\frac{1}{s}\sqrt{\frac{t}{u}}f(s,t,u)+2ie^2Q_{q}^2\sqrt{\frac{t}{u}}S(t,u) 
\label{eq:5}
\end{eqnarray}

\underline{(2). $q_{\alpha}\bar{q}_{\beta} \to \gamma_{\beta}\gamma_{\alpha};\  \alpha,\beta=L,R;\ \alpha\neq\beta=L,R$ }
\vskip 2mm
This is just $t \leftrightarrow u$ of the first case.  Note that because two photons are Abelian to each other, Chan-Paton parameters, $T$, in both cases are exactly the same.  With the same low-energy approximation as in 4-photon case, writing $A_{string}=A_{QED}+A_{cor}$, we have

\begin{eqnarray}
\bar{\Sigma}|A_{string}|^2 & = & \frac{1}{3}(\frac{1}{4})\left(\Sigma|A_{QED}|^2-2\Sigma|A_{QED}||A_{cor}|+\Sigma|A_{cor}|^2\right)
\label{eq:6}
\end{eqnarray}
where
\begin{eqnarray}
\Sigma|A_{cor}|^2 & = & \frac{e^4}{4}(\frac{\pi^4}{9})(Q_{q}^2+\frac{3}{2}T)^2(\frac{s^4}{M_{S}^8})(1-\cos^4{\theta})
\label{eq:7} \\
-2\Sigma|A_{QED}||A_{cor}| & = & -4e^4(\frac{\pi^2}{3})(Q_{q}^2+\frac{3}{2}T)Q_{q}^2(\frac{s^2}{M_{S}^4})(1+\cos^2{\theta})
\label{eq:8} \\
\Sigma|A_{QED}|^2 & = & 16e^4Q_{q}^4(\frac{1+\cos^2{\theta}}{1-\cos^2{\theta}})
\label{eq:9} 
\end{eqnarray}
with $u/s=-\frac{1}{2}(1+\cos{\theta}), t/s=-\frac{1}{2}(1-\cos{\theta})$.  We have also assume that all $T$'s are color-blind and they are the same for every pair of quark-antiquark.

Now we turn to $gg\rightarrow \gamma\gamma$.  Since the final state is color-neutral, the initial gluons must be a color singlet and therefore they must have opposite helicity.

\underline{(1). $g_{\alpha}g_{\beta} \rightarrow \gamma_{\alpha}\gamma_{\beta}$ with $\alpha, \beta = L, R, \alpha \neq \beta$.}  
\begin{eqnarray}
A_{string}=ig^2T\frac{t}{us}f(s,t,u)
\end{eqnarray}

\underline{(2). $g_{\alpha}g_{\beta} \rightarrow \gamma_{\beta}\gamma_{\alpha}$ with $\alpha, \beta = L, R, \alpha \neq \beta$.}  
\begin{eqnarray}
A_{string}=ig^2T\frac{u}{ts}f(s,t,u)(t\leftrightarrow u \mbox{ of above})
\end{eqnarray}

We assume that Chan-Paton factors are independent of the helicity of initial gluons i.e. $tr(t^{L}_{g}t^{L}_{g})=tr(t^{L}_{g}t^{R}_{g})=tr(t^{R}_{g}t^{R}_{g})$.  This is equivalent to the statement that the interaction is non-chiral.  In this non-chiral case, all Chan-Paton factors are the same $T$.  These amplitudes then give
\begin{eqnarray}
\bar{\Sigma}|A_{string}|^2 & = & \frac{g^4\pi^4T^2}{64}(\frac{s^4}{M_{S}^8})\frac{1}{8}(1+6\cos^2{\theta}+\cos^4{\theta})
\label{eq:10}
\end{eqnarray}
The question of which coupling $g$ should be identified with now arises.  In standard model, the interaction is achieved by tree-level graviton exchange($\sim 1/M_{Pl}^2$) and through gauge interactions at one-loop level($\sim \alpha\alpha_{s}$).  However, we are modelling stringy corrections to QED coupling and therefore we set $g=e$.  Strange as it looks from the viewpoint of field theory considering there is no corresponding vertices to begin with, this is allowed in a string theoretic framework if $T\neq 0$.  Scattering amplitude of this kind could be of the same order of magnitude as the 4-photon scattering.  According to our expressions above, they are $1/M_{S}^4$-suppressed in low energy limit and they become larger as we approach $M_S$.  At the Tevatron, these contributions from $gg \to \gamma\gamma$ would be of negligible size due to the low luminosity of gluons in the parton distribution functions of proton and antiproton~\cite{bhhm}.  

\subsection{Scattering of Dilepton into Diphoton}

This is the same as $q\bar{q}$ initial state with $Q_{q}^2=1$.  We will start off our comparison between open-string and Kaluza-Klein expressions by focussing on the electron-positron initial state.  We can see the similarity between open-string, Eq.~(\ref{eq:7}-\ref{eq:8}), and Kaluza-Klein cross-section formula, Eq.~(9) of ref.~\cite{kkg}.  By making the identification
\begin{eqnarray}
\frac{M^4_{D}}{F} & = & \frac{M^4_{S}}{-\frac{\pi^2}{3}\alpha(Q_{q}^2+\frac{3T}{2})}
\label{eq:11} \\
                  & = & \frac{\Lambda_{+}^4}{2\alpha}
\end{eqnarray}        
where $F=\log(M^2_D/s), 2/(n-2)$ for $n=2, n>2$ when $n$ is the number of extra dimensions.  $M_D$ is mass cutoff in Kaluza-Klein model\cite{kk}, and $\Lambda_{+}$ is the Drell's QED cutoff.  Following statistical analysis of Cheung's~\cite{kkg}, for example, at OPAL($\sqrt{s}=189$ GeV), $\Lambda_{+}>345$ GeV, we have
\begin{eqnarray}
M_{S} & > & 0.33-0.59 \mbox{ TeV for }T=(-1)-(-4).
\end{eqnarray}   
Note that the same data gives $M_{D}>0.98$ TeV for $n=4$(number of extra dimensions).  Considering the fact that $M_D$ is related to the quantum gravity scale which is larger than string scale, $M_S$, generically, this result is consistent.  The similarity between open-string and Kaluza-Klein amplitudes in this case is nevertheless not surprising as the first correction to the $e^{+}e^{-}\rightarrow \gamma \gamma$ is generated by a unique dimension-8 operator~\cite{pes}.  We therefore expect the same similarity in $q\bar{q}\rightarrow \gamma \gamma$ between Kaluza-Klein and open-string amplitudes and we will see that this is the case.  

\section{Comparison between open-string and Kaluza-Klein amplitudes}

Exactly the same identification, Eq.~(\ref{eq:11}) works in $q\bar{q}\rightarrow \gamma \gamma$ scattering.  To translate the statistical analysis of Cheung's\cite{kkg}, we assume $Q_{q}^2\approx 1/2$ to be the same for all quarks.  For Tevatron Run II, $n=4(M_{D}>1.43$ TeV)
\begin{eqnarray} 
M_{S} & \simeq & (0.40-0.61)M_{D} \mbox{ for } T=(-1)-(-4)\\
M_{S} & > & 0.57-0.87 \mbox{ TeV for } T=(-1)-(-4)  
\end{eqnarray}
This limit is consistent with the limit from dilepton production in \cite{bhhm}.  

Next we turn to the 4-boson cases, namely $gg\rightarrow \gamma\gamma$ and $\gamma\gamma \rightarrow \gamma\gamma$.  In $gg\rightarrow \gamma\gamma$, Eq.~(10) of ref.~\cite{kkg} has exactly the same energy-dependence and angular distribution as Eq.~(\ref{eq:10}) for the open-string case, the correspondence is 
\begin{eqnarray}
\frac{F}{M_{D}^4} & = & \pm \frac{\frac{\pi^2}{2}T\alpha}{M_{S}^4}
\label{eq:12}
\end{eqnarray}
For the 4-photon scattering, comparing Eq.~(\ref{eq:3}) with Eq.~(5) of \cite{kkg}, we have the correspondence
\begin{eqnarray}
\frac{F}{M_{D}^4} & = & \pm \frac{\frac{\pi^2}{4}T\alpha}{M_{S}^4}
\label{eq:13}
\end{eqnarray}
The scattering of 2 photons into 2 photons is forbidden at tree-level of QED.  The first non-zero QED-contribution comes from one-loop fermion scattering.  In open-string models, however, the tree-level QED-strength string-amplitudes are generically non-zero.  Assignment of a diagonal matrix to the photon is unique and the trace of the product of four matrices is non-vanishing in any $U(n)$ and therefore the amplitude does exists at tree-level.  The low value of these tree-level amplitudes, Eq.~(\ref{eq:1}-\ref{eq:2}), is a result of $s,t,u$ symmetry in open-string amplitude.  

In this aspect, 4-photon scattering is special; any open-string model gives a non-zero tree-level($\simeq \alpha$) amplitude for the process and the cross-section increases with energy.  Moreover, the background from $\alpha^2$-terms in QED is reduced as energy increases (Fig. 1 of \cite{kkg}) in the energy range we can probe in current and future colliders.  Limits on lower bound of string scale $M_S$ obtained from 4-photon scattering would be universal for every open-string model and it would be used as the standard normalization for the value of Chan-Paton parameters.  

\section{Understanding string-KK Similarity in Diphoton production}

The open-string amplitudes we use are the formula for scattering of 4 particles through gauge bosons exchange extended to string scattering by Veneziano extension(\ie~ multiplying the corresponding channel by $S(s,t), S(t,u)$ and $S(u,s)$).  For massless external particles, the amplitudes are factorized into three distinct helicity combinations.  Each one approaches different field-theory limits in low energy corresponding to $1/s,1/t$ and $1/u$ gauge-boson propagators\cite{man}.  In other words, the $0$th resonance of the formula corresponds to gauge-boson(spin-1) exchange in field theory expansion.  String correction comes in as exchange of higher spin states as we see from
\begin{eqnarray}
S(s,t) & \simeq & 1-\frac{\pi^2}{6}(\frac{st}{M_{S}^4})
\end{eqnarray} 
where string correction is the 2nd term in the expression.  Additional power of $t$ in the numerator of the correction term brings an additional spin state to the intermediate state.  In this case, there is spin-2 exchange in addition to the spin-1(gauge boson) exchange due to the correction term.  Notably, $S(t,u)$ brings in 1 and 2 additional spins since $u=-s-t$ and $ut=-st-t^2$ and we will end up with spin-1,2 and 3 exchange in the amplitude containing single $S(t,u)$ term.    

From the above general argument, we analyze the diphoton production processes.  To illustrate important point, we first consider the 4-photon scattering.  As we see from Eq.~(\ref{eq:1}-{\ref{eq:2}), at low energy, the first-order amplitudes are proportional to $s/t, s/u$ and $u/s, u/t$(they add up to zero in each helicity case at the leading order before string corrections) and therefore appear to have spin-1 exchange contributions.  However Yang's theorem\cite{yan} implies that these contributions actually are spin-0 components of gauge boson exchange as we can see easily from explicitly writing down tree-level Feynman diagrams.  Therefore the string corrections add spin exchange up to spin-2 for each helicity combination, proportional to $s^2, t^2$ and $u^2$ respectively.  After the corrections, Yang's theorem again prevents spin-1 exchange after the string corrections and we are left with only spin-2 exchange(since the original spin-0 scattering vanishes without string corrections).  

The same argument applies to other processes with diphoton final state, they start with only spin-0 exchange before string corrections and end up with spin-2 exchange and original SM-exchange(processes such as $q\bar{q}, e^{+}e^{-}\to \gamma\gamma$ have non-vanishing SM part that remains) after string corrections.  Low-energy open-string corrections of amplitudes for diphoton production thus are spin-2 exchange in nature.  On the other hand, KK corrections in conventional KK model\cite{kk} are naturally spin-2 exchange.  It was shown by Feynman, Kraichnan, and Weinberg\cite{feyn} that Lorentz-invariant CPT-preserved spin-2 exchange interaction is unique and similarity between open-string and Kaluza-Klein in diphoton production is a manifestation of this.  The only difference between them is the strengths of couplings.  Gravitational(KK) is much weaker than SR-extended QED interaction due to different spaces of propagation.  Summation over KK tower amplifies KK contribution to somewhat the same scale as SR contribution.  In some regions of parameter space, we therefore expect to have two copies of dimension-8 operators correcting standard model amplitudes.  The scale could be as low as 1 TeV due to previous estimations\cite{pes,bhhm} and our current results.            

\section{Conclusions}
\label{conclusions-sec}

Tree-level open-string scattering amplitudes of various diphoton production processes have been calculated with unspecified Chan-Paton parameters.  With the assumption of nonchiral interaction, we found remarkably similar forms(for both energy dependence and angular distribution) between low energy open-string scattering amplitudes and diphoton production amplitudes via graviton exchange in Kaluza-Klein model regardless of the fact that one is confined to D3brane and the other propagates freely in the bulk.  Applying the low energy constraints on mass scale of one model to another is therefore allowed and we extract some constraints on string scale $M_S$ from constraints on $M_D$ in KK model.  We found an agreement(somewhat weaker) with other constraints on $M_S$ from the 4-fermion processes\cite{bhhm}, about $0.6-0.9$ TeV.  Also we emphasize that the 4-photon interaction is unique and universal in every open-string model as well as it is phenomenologically clean from SM background.  

Caution thus has to be made when doing new physics analysis on diphoton production.  There could be two copies of exactly the same form of corrections to SM amplitudes, one from SR(low $E$) and one from KK(tower of states).  In some cases of low-scale string scenario with large compatified extra dimension, the SR corrections to SM amplitudes will be dominant and they will show up first since KK-graviton exchange is more suppressed by higher power of coupling\cite{pes}.  KK-graviton exchange will also show up as smaller contributions with exactly the same angular distribution and energy dependence at a somewhat higher scale($M_D>M_S$).  However, it is also possible that the string scenario is of much higher scale or not valid at all, in which case we might find only one copy of corrections to SM diphoton amplitudes coming from conventional field theoretic KK exchange.  In the intermediate kinematic region($100$ GeV $<E<M_S, M_D$), we need cross-check from other channels like the 4-fermion scattering to distinguish between signals from SR and signals from KK models.  As we go to higher energy($E>M_S$), since we can investigate the resonances directly, the detailed energy and angular distributions at the resonances will determine whether it is SR or KK exchange with more certainty\cite{bfh}.

\section*{Acknowledgments}

\indent
I would like to thank Tao Han for previous collaboration from which many results have been used in this paper as well as valuable  and insightful discussions and Keith Thomas for discussions and carefully reading through and correcting the manuscript.  This work was supported in part by the U.S. Department of Energy 
under grant number DE-FG02-95ER40896, and in part by the Wisconsin
Alumni Research Foundation.

\appendix 

\section{String Amplitudes}

General form for scattering amplitude for 4 external massless particles in conventional TeV-string scenario is given in ref.~\cite{pes, has} as
\begin{eqnarray}
A(s,t,u) & = & ig_s \left[T_{1234}A_{1234}S(s,t)+T_{1324}A_{1324}S(t,u)+T_{1243}A_{1243}S(u,s) \right]
\end{eqnarray}
where $T_{ijkl}=tr(t^{i}t^{j}t^{k}t^{l})+tr(reverse)$ are Chan-Paton factors for matrices $t^{i}$ of $U(n)$ gauge group with normalization $tr(t^{i}t^{j})=\delta^{ij}$.  $A_{ijkl}$ are kinematic amplitudes listed below.  $S(x,y)$ is closely related to the Veneziano amplitude, and is defined by
\begin{eqnarray}
S(x,y) & = & \frac{\Gamma(1-\alpha^{\prime}x)\Gamma(1-\alpha^{\prime}y)}{\Gamma(1-\alpha^{\prime}x-\alpha^{\prime}y)},
\end{eqnarray}
where the Regge slope parameter $\alpha^{\prime}=M^{-2}_S$.  $S\to 1$ as $M_S \gg \sqrt{s}$.  $g_s=g^2$ is the string coupling which will be identified as gauge coupling $g^2_{YM}$ in our setup.  
    
\section{Kinematic Table}

This is the table we use to calculate kinematic parts of tree-level amplitudes, $f(g)$ represents fermion(boson).  

\begin{tabular}{|c|c|c|}
\hline
   & gg & ff \\ \hline
  gg & G & E \\ \hline
  ff & F & G \\ \hline
\end{tabular} \\

$G = A_{1234},A_{1324},A_{1243}(gggg, ffff)$

$E = A_{1234},A_{1324},A_{1243}(ggff)$

$F = A_{1234},A_{1324},A_{1243}(ffgg)$

\begin{eqnarray}
  (g^{-}g^{+}g^{-}g^{+});  A_{1234} &=& ig^2 \frac{\langle13\rangle^4}{\langle12\rangle^2 \langle14\rangle^2} \nonumber \\
                           A_{1324} &=& ig^2 \frac{\langle13\rangle^2}{\langle14\rangle^2}\nonumber \\
                           A_{1243} &=& ig^2 \frac{\langle13\rangle^2}{\langle12\rangle^2}\nonumber
\end{eqnarray}

\begin{eqnarray}
  (g^{-}g^{-}g^{+}g^{+});  A_{1234} &=& ig^2 \frac{\langle12\rangle^2}{\langle14\rangle^2} \nonumber \\
                           A_{1324} &=& ig^2 \frac{\langle12\rangle^4}{\langle13\rangle^2 \langle14\rangle^2}\nonumber \\
                           A_{1243} &=& ig^2 \frac{\langle12\rangle^2}{\langle13\rangle^2}\nonumber
\end{eqnarray}

\begin{eqnarray}
  (g^{-}g^{+}g^{+}g^{-});  A_{1234} &=& ig^2 \frac{\langle14\rangle^2}{\langle12\rangle^2} \nonumber \\
                           A_{1324} &=& ig^2 \frac{\langle14\rangle^2}{\langle13\rangle^2}\nonumber \\
                           A_{1243} &=& ig^2 \frac{\langle14\rangle^4}{\langle12\rangle^2 \langle13\rangle^2}\nonumber
\end{eqnarray}

\begin{eqnarray}
  (g^{-}g^{+}f^{+}f^{-});  A_{1234} &=& ig^2 \frac{\langle13\rangle \langle14\rangle}{\langle12\rangle^2} \nonumber \\
                           A_{1324} &=& ig^2 \frac{\langle14\rangle}{\langle13\rangle}\nonumber \\
                           A_{1243} &=& ig^2 \frac{\langle14\rangle^3}{\langle13\rangle \langle12\rangle^2}\nonumber
\end{eqnarray}

\begin{eqnarray}
  (f^{+}f^{-}g^{+}g^{-});  A_{1234} &=& ig^2 \frac{\langle13\rangle^3}{\langle12\rangle^2 \langle14\rangle} \nonumber \\
                           A_{1324} &=& ig^2 \frac{\langle13\rangle}{\langle14\rangle}\nonumber \\
                           A_{1243} &=& ig^2 \frac{\langle13\rangle \langle14\rangle}{\langle12\rangle^2}\nonumber
\end{eqnarray}

\begin{eqnarray}
  (f^{-}g^{+}g^{-}f^{+});  A_{1234} &=& ig^2 \frac{\langle13\rangle^3}{\langle12\rangle \langle14\rangle^2} \nonumber \\
                           A_{1324} &=& ig^2 \frac{\langle12\rangle \langle13\rangle}{\langle14\rangle^2}\nonumber \\
                           A_{1243} &=& ig^2 \frac{\langle13\rangle}{\langle12\rangle}\nonumber
\end{eqnarray}

\begin{eqnarray}
  (f^{-}g^{-}g^{+}f^{+});  A_{1234} &=& ig^2 \frac{\langle12\rangle \langle13\rangle}{\langle14\rangle^2} \nonumber \\
                           A_{1324} &=& ig^2 \frac{\langle12\rangle^3}{\langle13\rangle \langle14\rangle^2}\nonumber \\
                           A_{1243} &=& ig^2 \frac{\langle12\rangle}{\langle13\rangle}\nonumber
\end{eqnarray}
All momenta are incoming.  These were derived from the following general
formula \cite{man};
\begin{eqnarray}
A_{1234} = ig^2 \frac{\langle IJ \rangle^4}{\langle12\rangle
\langle23\rangle \langle34\rangle \langle41\rangle}\
\end{eqnarray}

I, J are momenta with negative helicity.  Order of $\langle XY
\rangle$ in the denominator is cyclic of 1234.  And also equation
(4.9) of ref.~\cite{man} has been applied for fermions involving
processes.

}


\end{document}